\documentclass[10pt,aps,pre,twocolumn,superscriptaddress,showpacs]{revtex4-1}

\usepackage{enumerate,amssymb,ascmac,amsmath,bm}
\usepackage[dvipdfmx]{graphicx,color}
\newcommand{\dis}{\displaystyle} 
	
\usepackage[dvipdfmx]{color}

\begin{document}

	\title{Irreversible simulated tempering}
	
	\author{Yuji Sakai}
	\email[]{yuji0920@huku.c.u-tokyo.ac.jp}
	\affiliation{Graduate School of Arts and Sciences, The University of Tokyo, 
	3-8-1 Komaba, Meguro-ku, Tokyo 153-8902, Japan}
	
	\author{Koji Hukushima}
	\email[]{hukusima@phys.c.u-tokyo.ac.jp}
	\affiliation{Graduate School of Arts and Sciences, The University of Tokyo, 
	3-8-1 Komaba, Meguro-ku, Tokyo 153-8902, Japan}
	\affiliation{Center for Materials Research by Information Integration,
	National Institute for Materials Science, 1-2-1 Sengen, Tsukuba, 
	Ibaraki 305-0047, Japan}
	
	\date{\today}
	
		\begin{abstract}
		An extended ensemble Monte Carlo algorithm is proposed 
		by introducing a violation of the detailed balance condition 
		to the update scheme of the inverse temperature in simulated tempering.
		Our method, irreversible simulated tempering, is constructed 
		based on the framework of the skew detailed balance condition.
		By applying this method to the ferromagnetic Ising model in two dimensions 
		on a square lattice as a benchmark, 
		the dynamical behavior of the inverse temperature and 
		an autocorrelation function of the magnetization are studied numerically.
		It is found that the relaxation dynamics of the inverse temperature 
		changes qualitatively from diffusive to ballistic 
		by violating the detailed balance condition. 
		Consequently, the autocorrelation time of magnetization is 
		several times smaller than that for the conventional algorithm 
		satisfying the detailed balance condition. 
		\end{abstract}

	
	\maketitle


	\section{\label{Sect:Introduction} Introduction}
		
	Since Metropolis et al.\ invented a Markov-chain Monte Carlo (MCMC) method 
	in 1953~\cite{Metropolis1953}, it has been widely implemented 
	in various research fields to evaluate expectation values for 
	a high-dimensional probability distribution.
	Meanwhile, some improvement and development of the MCMC method 
	have been made for more efficient sampling. 
	Among them, simulated tempering~\cite{Lyubartsev1992, Marinari1992}, 
	developed in the field of statistical physics, is categorized as 
	an extended ensemble method. 
	In simulated tempering, the inverse temperature 
	in the Gibbs-Boltzmann distribution is treated as a random variable 
	as well as the configurations and thus the state space of the system is 
	extended by adding the temperature.
	A Markov chain on the extended state space is constructed with 
	a detailed balance condition (DBC), which is a sufficient condition for MCMC. 
	The simulated tempering has been eagerly used for various problems in 
	statistical physics~\cite{Vicari1993, Kerler1994, Coluzzi1995, Fernandez1995, Picco1998} and protein-folding problems~\cite{Irback1995, Irback1997}. 
	
	Recently, a lifting technique in which the detailed balance condition (DBC) 
	in the Markov chain is broken with the global balance condition 
	still holding has been extensively studied.  
	Several studies have shown that diffusive relaxation dynamics in 
	a one-dimensional random walk with DBC is qualitatively improved by using 
	the lifting technique~\cite{Chen1999, Diaconis2000, YS2015a, Vucelja2014}.	
	The transition graph of the inverse temperature under a fixed configuration 
	in the simulated tempering is the same as that in the random walk.
	Thus, we expect that the violation of DBC makes the relaxation dynamics of 
	the inverse temperature in simulated tempering change qualitatively. 
	
	In this study, we propose to apply the idea of 
	a skew detailed balance condition (SDBC)~\cite{YS2015a,Turitsyn2011} 
	to the update scheme of the inverse temperature in simulated tempering, 
	thus conducting the simulated tempering algorithm without DBC. 	
	As a benchmark, we examine the efficiency of our proposed algorithm 
	in a two-dimensional Ising model and show, numerically, that SDBC changes, 
	qualitatively, the relaxation dynamics of the inverse temperature 
	in simulated tempering. 
	Furthermore, we observe that the autocorrelation time of 
	the magnetization is also reduced by the violation of DBC.
	
	This paper is organized as follows. Sec.~\ref{Sect:ST} introduces 
	the simulated tempering method satisfying DBC. 
	In Sec.~\ref{Sect:STwithSDBC}, a simulated tempering algorithm with SDBC 
	is constructed. 
	We apply the proposed algorithm to an Ising model in two dimensions as 
	a benchmark and confirm the efficiency of our proposed algorithm 
	in Sec.~\ref{Sect:Benchmark}. 
	Section~\ref{Sect:Summary} summarizes the present work. 

	
	\section{\label{Sect:ST}Simulated tempering}
		
	In this section, we explain the outline of 
	simulated tempering~\cite{Lyubartsev1992, Marinari1992} 
	in order to fix our notation.	
	
	\subsection{\label{subsect:setup}Setup}	
	
	Let $\bm{X}$ be a configuration to be sampled from 
	a target distribution function in MCMC simulations. 
	In statistical physics, the target distribution	$P_{\beta}(\bm{X})$ 
	is often given by the Gibbs-Boltzmann distribution with 
	an inverse temperature $\beta$,	
		\begin{align}
		P_{\beta}(\bm{X})=\frac1{Z(\beta)}\exp[-\beta E(\bm{X})],
		\label{eq:P_beta}
		\end{align}
	where $E(\bm{X})$ is a model Hamiltonian and $Z(\beta)$ is 
	the partition function of the model.		
	In simulated tempering~\cite{Lyubartsev1992,Marinari1992}, 
	the (inverse) temperature, as well as the configuration, 
	is a random variable. 
	More specifically, $\beta$ takes $R$ different values determined 
	before the simulation, expressed as $\beta_1,\ldots,\beta_R$.
	Thus, a state is specified by these variables, denoted by $(\bm{X},\beta_r)$. 
	Then, an extended equilibrium distribution $P_{\rm ST}(\bm{X},\beta_r)$ 
	for finding a state $(\bm{X},\beta_r)$ is given as 
		\begin{align}
		P_{\rm ST}(\bm{X},\beta_r)=\frac1{Z_{\rm ST}}
		\exp[-\beta_r E(\bm{X})+g_r],
		\label{eq:P_ST}
		\end{align}
	where the weight factor $g_r$ is a constant depending only on 
	the inverse temperature and the extended partition function $Z_{\rm ST}$ is
		\begin{align}
		Z_{\rm ST}=
		\sum_{r=1}^{R}\sum_{\bm{X}}
		\exp[-\beta_r E(\bm{X})+g_r]=
		\sum_{r=1}^{R}Z(\beta_r)e^{g_r}.
		\end{align}
	For a given $\beta_r$, the probability for finding a configuration 
	$\bm{X}$ in Eq.~(\ref{eq:P_ST}) is proportional to that 
	in Eq.~(\ref{eq:P_beta}). 
	The average over the sampled configurations conditioned with $\beta_r$ 
	is equivalent to the equilibrium average at that temperature. 
	In contrast, the marginal probability for a given $\beta_r$ is obtained by 
		\begin{align}
		P_{\mathrm{ST}}(\beta_r)=
		\sum_{\bm{X}}P_{\mathrm{ST}}(\bm{X},\beta_r)=
		\frac{Z(\beta_r)e^{g_r}}{Z_{\rm ST}}.
		\label{eq:marginal-P-beta}
		\end{align}
	The marginal probability is independent of $r$ when	$g_r=-\ln{Z(\beta_r)}$, 
	which is proportional to the bulk free energy at $\beta_r$ 
	of the model simulated.  
	In general, it is hard to estimate the value of the free energy 
	for a statistical-mechanical model. 
	However, if it could be estimated a priori, even approximately, 
	a uniform sampling for $\beta$ from high to low temperatures can be 
	put into practice.
	By considering an appropriate Markov chain, the state of $\beta_r$ wanders 
	on the temperature axis in a random-walk manner. 
	One may expect that it is relatively easy to sample	configurations 
	at sufficiently high temperatures, which would help an efficient sampling 
	at low temperatures through the random walk of $\beta_r$. 
	This is what we expect to perform in simulated tempering.

		
	\subsection{Simulated tempering algorithm with detailed balance condition}
	
	An explicit update scheme of the simulated tempering method consists of 
	the following two steps: an update of a configuration $\bm{X}$ for 
	a fixed $\beta_r$ and an update of $\beta_r$ for a fixed $\bm{X}$. 
	In order to generate a Markov chain, the corresponding two 
	transition-probability matrices are introduced. 
	One is the transition matrix from a state $(\bm{X},\beta_r)$ to 
	$(\bm{X}',\beta_r)$ denoted as $T(\bm{X}',\beta_r|\bm{X},\beta_r)$.
	The other is one from $(\bm{X},\beta_r)$ to $(\bm{X},\beta_l)$ as 
	$T(\bm{X},\beta_l|\bm{X},\beta_r)$.
	They satisfy DBC for the stationary distribution of Eq.~(\ref{eq:P_ST}). 
	In practice, the Metropolis-Hastings type~\cite{Hastings1970} of 
	the transition probabilities is often used.
	Here, we assume that the transition probabilities are decomposed into
		\begin{align}
		&
		T(\bm{X}',\beta_r|\bm{X},\beta_r)
		\notag \\
		&~=
		q(\bm{X}',\beta_r|\bm{X},\beta_r)
		W(\bm{X}',\beta_r|\bm{X},\beta_r),
		\label{eq:T1}
		\end{align}
	and
		\begin{align}
		&
		T(\bm{X},\beta_l|\bm{X},\beta_r)
		\notag \\
		&~=
		q(\bm{X},\beta_l|\bm{X},\beta_r)
		W(\bm{X},\beta_l|\bm{X},\beta_r),
		\label{eq:T2}
		\end{align}
	where $q$ denotes the proposal probability and $W$ is 
	the acceptance probability.
	Then, the explicit forms of the Metropolis-Hastings type of 
	the acceptance probabilities for Eq.~(\ref{eq:T1}) and (\ref{eq:T2}) 
	are given by
		\begin{align}
		&
		W(\bm{X}',\beta_r|\bm{X},\beta_r)
		\notag \\
		&~=
		\min\left[1,
		\frac{q(\bm{X},\beta_r|\bm{X}',\beta_r)}{
		q(\bm{X}',\beta_r|\bm{X},\beta_r)}
		\frac{P_{\rm ST}(\bm{X}',\beta_r)}{P_{\rm ST}(\bm{X},\beta_r)}\right],
		\end{align}
	and 
		\begin{align}
		&
		W(\bm{X},\beta_l|\bm{X},\beta_r)
		\notag \\
		&~=
		\min\left[1,
		\frac{q(\bm{X},\beta_r|\bm{X},\beta_l)}{q(\bm{X},\beta_l|\bm{X},\beta_r)}
		\frac{P_{\rm ST}(\bm{X},\beta_l)}{P_{\rm ST}(\bm{X},\beta_r)}\right],		
		\end{align}
	respectively.
	
\begin{figure*}[t]
		\centering
		\includegraphics[width=.95\textwidth,clip]{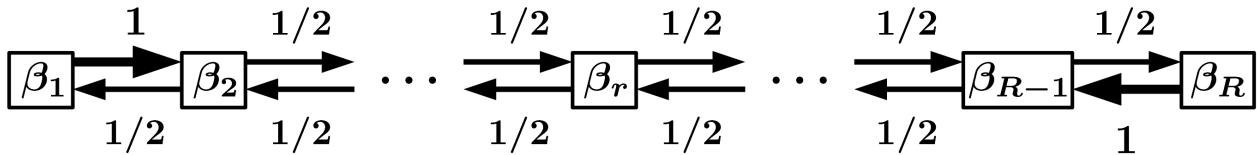}
		\caption{Graphical expression of the proposal probabilities $q_{r,l}$ 
		in  conventional simulated tempering.}
		\label{fig:proposal-GT}
		\end{figure*}

	For simplicity, the set of inverse temperatures are ordered 
	such that $\beta_1<\beta_2<\cdots<\beta_R$. 
	In addition, throughout this paper we use the proposal probability 
	$q_{r,l}\equiv q(\bm{X},\beta_l|\bm{X},\beta_r)$ given by 
	$q_{1,2}=q_{R,R-1}=1$ and $q_{r,r\pm 1}=1/2$ if $1<r<R$, 
	and zero otherwise (Fig.~\ref{fig:proposal-GT}). 
	Note that $q_{r,l}$ is independent of the configuration $\bm{X}$.
	Then, the procedure of the simulated tempering method is 
	described as follows.
		\begin{enumerate}
		\item Arbitrarily chose an initial state $(\bm{X}^{(0)},\beta^{(0)})$.
		\item Iterate the update trials for an original configuration $\bm{X}$ 
		at a fixed $\beta$ according to the conventional 
		Metropolis-Hastings method:  
			\begin{enumerate}
			\item Suppose that the current state is $(\bm{X},\beta_r)$ and 
			select a configuration $\bm{X}'$ with the probability 
			$q(\bm{X}',\beta_r|\bm{X},\beta_r)$.
			\item Accept the new state $(\bm{X}',\beta_r)$ with the probability 
			$W(\bm{X}',\beta_r|\bm{X},\beta_r)$. 
			If it is rejected, set the current state as the next state.
			\end{enumerate}
		\item Iterate the update trials for an inverse temperature $\beta$ 
		according to the following procedure:
			\begin{enumerate}
			\item Suppose that the current state is $(\bm{X},\beta_r)$ and 
			choose $\beta_{l}$ with the probability $q_{r,l}$ 
			as a candidate for the next inverse temperature 
			(Fig.~\ref{fig:proposal-GT}).
			\item Accept the next state $(\bm{X},\beta_l)$ with the probability 
			$W(\bm{X},\beta_l|\bm{X},\beta_r)$. 
			If it is rejected, set the current state as the next state.
			\end{enumerate}
		\end{enumerate}
	Although any inverse temperature could be chosen as a candidate in step 3, 
	in practice we only consider the nearest ones 
	to increase the transition rate.
	By repeating steps 2 and 3, our desired Markov chain of the state 
	$(\bm{X},\beta)$ is obtained.
	
	Since the simulated tempering was invented, it has been widely applied 
	to various problems~\cite{Vicari1993, Kerler1994, Coluzzi1995, Fernandez1995, Picco1998, Irback1995, Irback1997}. 
	In addition, the improvement of simulated tempering has been 
	continuously studied~\cite{Geyer1995, Mitsutake2000, Mitsutake2004, Li2004, Nguyen2013, Mori2015}. 
	In particular, an efficient simulated tempering in which DBC breaks 
	and the rejection rate is decreased with the violation of DBC 
	using the Suwa-Todo algorithm~\cite{Suwa2010} was proposed recently 
	in Ref.~\cite{Mori2015}. 
	Others authors keep DBC unbroken. 
	The typical relaxation time of the algorithm is reduced by some factor, 
	but its temperature dynamics does not change quantitatively 
	from standard diffusive dynamics. 
	

	\section{\label{Sect:STwithSDBC}
	Simulated tempering with skew detailed balance conditions}
	
	In the conventional simulated tempering method explained 
	in the previous section, the transition graph of the inverse temperature 
	$\beta$ under fixed $\bm{X}$ in Fig.~\ref{fig:proposal-GT} is 
	the same as that of a one-dimensional simple random walk.
	Thus, $\beta$ behaves as a random walker on the graph.
	It is known that a random walk satisfying DBC is essentially 
	a diffusive process and its relaxation time is of order $O(\Omega^2)$, 
	where $\Omega$ denotes the number of states in the random walk. 
	Several studies have shown, numerically and analytically, that 
	the $\Omega$-dependence of the relaxation time is improved 
	by introducing the ``lifting'' technique to the random walk 
	in one dimension~\cite{YS2015a,Chen1999,Diaconis2000,Vucelja2014}. 
	Here DBC is broken by adding a lifting parameter 
	while preserving the global balance condition. 
	This strongly suggests that the performance of simulated tempering is 
	improved qualitatively by the lifting of the update of $\beta$. 
	In this section, we apply the methodology of 
	SDBC~\cite{YS2015a, Turitsyn2011} to simulated tempering, 
	especially to the update scheme of the inverse temperature.
	The proposed method is called irreversible simulated tempering. 

	
	\subsection{\label{subsect:setup-iST}Setup}
	
	Let us reconsider the setup of conventional simulated tempering 
	to extend it to an irreversible one.
	By introducing an auxiliary random variable $\varepsilon\in\{+,-\}$ 
	to the system as a lifting parameter, the state space is duplicated.
	A state in the duplicated state space is denoted 
	$(\bm{X},\beta_r,\varepsilon)$. 
	Then, the extended equilibrium distribution 
	$P_{\rm ST}(\bm{X},\beta_r,\varepsilon)$ for finding a state 
	$(\bm{X},\beta_r,\varepsilon)$ is given by
		\begin{align}
		P_{\rm ST}(\bm{X},\beta_r,\varepsilon)=
		\frac1{2Z_{\rm ST}}\exp[-\beta_r E(\bm{X})+g_r].
		\label{eq:extendedP_{ST}}
		\end{align}
	Note that the marginal probability for a given configuration 
	$\bm{X}$ and an inverse temperature $\beta_r$ is exactly the same as 
	$P_{\rm ST}(\bm{X},\beta_r)$ and that it is uniform 
	for a given $\varepsilon$.
	
\begin{figure*}[t]
		\centering
		\includegraphics[width=.95\textwidth,clip]{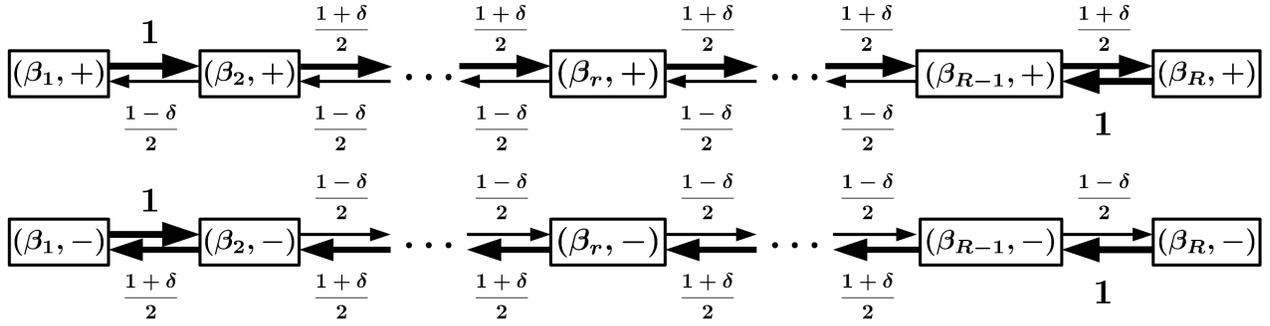}
		\caption{Graphical expression of the proposal probabilities 
		$q_{r,l}^{(\varepsilon)}$ in an irreversible simulated tempering.}
		\label{fig:proposal-GT-2}
		\end{figure*}

	In this study, we apply the methodology of SDBC only 
	to the update scheme of the inverse temperature. 
	The skew detailed balance condition in this context is expressed as
		\begin{align}
		&T(\bm{X},\beta_l,+|\bm{X},\beta_r,+)
		P_{\rm ST}(\bm{X},\beta_r,+)\notag \\
		&~= 
		T(\bm{X},\beta_r,-|\bm{X},\beta_l,-)
		P_{\rm ST}(\bm{X},\beta_l,-),
		\end{align}
	where $T(\bm{X},\beta_l,\varepsilon|\bm{X},\beta_r,\varepsilon)$ 
	denotes the transition probability from a state 
	$(\bm{X},\beta_r,\varepsilon)$ to $(\bm{X},\beta_l,\varepsilon)$.
	Again, we decompose the transition probability 
	$T(\bm{X},\beta_l,\varepsilon|\bm{X},\beta_r,\varepsilon)$ 
	into the product of a proposal probability and 
	an acceptance probability expressed as
		\begin{align}
		&T(\bm{X},\beta_l,\varepsilon|\bm{X},\beta_r,\varepsilon)
		\notag \\
		&~=
		q(\bm{X},\beta_l,\varepsilon|\bm{X},\beta_r,\varepsilon)
		W(\bm{X},\beta_l,\varepsilon|\bm{X},\beta_r,\varepsilon). 
		\end{align}
	By using the proposal probability 
	$q(\bm{X},\beta_l,\varepsilon|\bm{X},\beta_r,\varepsilon)$, 
	the general form of the Metropolis-Hastings-type acceptance probability 
	that satisfies SDBC is explicitly given by
		\begin{align}
		&W(\bm{X},\beta_l,\varepsilon|\bm{X},\beta_r,\varepsilon)
		\notag \\
		&~=\min\left[1,
		\frac{q(\bm{X},\beta_r,-\varepsilon|\bm{X},\beta_l,-\varepsilon)}{
		q(\bm{X},\beta_l,\varepsilon|\bm{X},\beta_r,\varepsilon)}
		\frac{P_{\rm ST}(\bm{X},\beta_l)}{P_{\rm ST}(\bm{X},\beta_r)}\right],
		\end{align}
	In this study, we construct the proposal probability as 
	$q_{r,l}^{(\varepsilon)}$, which is independent of the configuration 
	$\bm{X}$ as follows (Fig.~\ref{fig:proposal-GT-2}):
		\begin{align}
		&q_{1,2}^{(\varepsilon)}=q_{R,R-1}^{(\varepsilon)}=1,\\
		&q_{r,r\pm1}^{(\varepsilon)}=\frac{1\pm\delta\varepsilon}2,
		\end{align}
	if $1<r<R$, and $q_{r,l}^{(\varepsilon)}=0$ otherwise.
	The parameter $\delta$ controls the violation of DBC and 
	satisfies $|\delta|<1$.
	DBC is restored when the parameter $\delta$ is set to zero.
	One expects that a finite positive value of $\delta$ enhances 
	a	clockwise flow in the dynamics of $\beta$ in Fig.~\ref{fig:proposal-GT-2}. 

						
	\subsection{\label{subsect:STwithSDBC}
	Simulated tempering algorithm with skew detailed balance conditions}
	
	For simplicity, we rewrite the acceptance probability 
	$W(\bm{X},\beta_l,\varepsilon|\bm{X},\beta_r,\varepsilon)$ 
	as $W_{r,l}^{(\varepsilon)}$. 
	Then, the update scheme of an inverse temperature $\beta$ and 
	an auxiliary variable $\varepsilon$ is described as follows:
		\begin{enumerate}
		\item  Iterate the update trials for an inverse temperature $\beta$	
		according to the following procedure:
			\begin{enumerate}
			\item Suppose that the current state is $(\bm{X},\beta_r,\varepsilon)$ 
			and choose $\beta_{l}$ with the probability $q_{r,l}^{(\varepsilon)}$ 
			as a candidate of the next inverse temperature.
			\item Accept the next state $(\bm{X},\beta_l,\varepsilon)$ with 
			the probability $W_{r,l}^{(\varepsilon)}$. 
			\item If the trial is rejected, flip $\varepsilon$ and set 
			$(\bm{X},\beta_r,-\varepsilon)$ as the next state with the probability 
				\begin{align}
				\lambda_r^{(\varepsilon)}\equiv
				\frac{\Lambda_r^{(\varepsilon)}}{\dis
				1-\sum_{l\not=r}q_{r,l}^{(\varepsilon)}
				W_{r,l}^{(\varepsilon)}},
				\end{align}
			where
				\begin{align}
				\Lambda_r^{(\varepsilon)}=
				\max\left[0,~-\varepsilon\sum_{\varepsilon'=\pm}
				\sum_{l\not=r}\varepsilon'
				q_{r,l}^{(-\varepsilon')}
				W_{r,l}^{(-\varepsilon')}
				\right].
				\end{align}
			If it is also rejected, set the current state as the next state.
			\end{enumerate}
	 	\item Update the configuration $\bm{X}$ with the conventional 
	 	Metropolis-Hastings method at the inverse temperature $\beta_l$, 
	 	as explained in Sec.~\ref{Sect:ST}. 
		\end{enumerate}
	It is straightforward to verify that the global balance condition is 
	satisfied in the above procedure~\cite{Turitsyn2011,YS2015a}.

	
	\section{\label{Sect:Benchmark}Benchmark}
	
	In this section, we apply the proposed algorithm to an Ising model 
	in two dimensions on a square lattice as a benchmark and 
	numerically evaluate the efficiency of the algorithm.
	Let $L$ denotes the linear size of the Ising model and let $N=L^2$.
	The energy function of the model is given by
		\begin{align}
		E(\bm{S})=-\sum_{\langle ij \rangle} S_i S_j,
		\end{align}
	with $\bm{S}=\{S_i\}_{i=1}^{N}$ and $S_i=\pm 1$.
	The sum is taken over the nearest neighbor pairs, 
	imposing a periodic boundary condition.
	The energy unit is set to unity. 
	The set of inverse temperatures in the simulated tempering is 
	prepared from $\beta_1=0.2$ and $\beta_R=0.5$, 
	with the intermediate values equally spaced between them.
	Note that the critical inverse temperature of the model is known as	
	$\beta_{\rm c}=\ln(1+\sqrt{2})/2\simeq 0.4407$~\cite{Onsager1944}, 
	which is inside the temperature region in our simulations. 
	
	In the present work, we focus our attention on the performance of 
	irreversible simulated tempering for an ideal weight parameter $g$. 
	Thus, the parameter $g_r$, which should be given a priori, is 
	evaluated by using an exact free energy numerical method with 
	a polynomial time of $N$~\cite{Kastening2001} available 
	for finite-size Ising models in two dimensions.
	The determination of the parameter, which is an important issue 
	when actually applying the algorithm to some statistical-mechanics models, 
	is discussed in a separate paper~\cite{YS2016b}. 

			
	\subsection{Relaxation dynamics of the inverse temperature}
	
\begin{figure}[t]
		\centering
		\includegraphics[width=.95\columnwidth,clip]{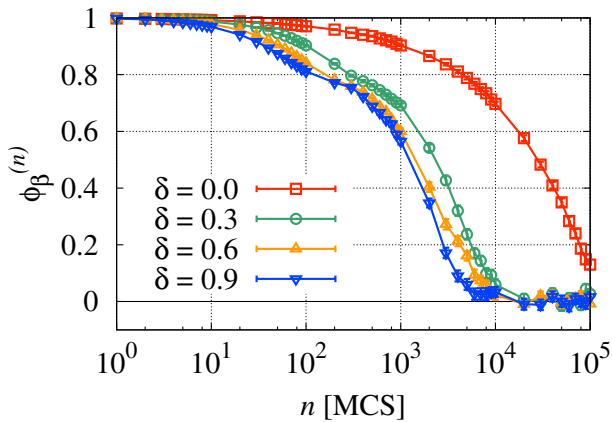}
		\caption{(Color online)
		Time evolution of the relaxation function of the inverse temperature 
		$\beta$ in simulated tempering with SDBC, applied to the Ising model 
		in two dimensions with $L=2^5$. 
		The initial state $\bm{S}^{(0)}$ is prepared by 
		the Metropolis-Hastings algorithm at $\beta=\beta_R=0.5$.
		The number of observed temperatures is $R=2^9$.
		The parameter $\delta$, which characterizes the deviation from DBC, 
		is chosen as $\delta=0.0$ (red square), $\delta=0.3$ (green circle), 
		$\delta=0.6$ (orange triangle), and 
		$\delta=0.9$ (blue inverted triangle), respectively.
		The history average is taken over $2^{10}$ samples and 
		the error bars are of the order of the symbol sizes. 
		}
		\label{fig:plot-beta}
		\end{figure}

	We study the dynamics of the inverse temperature in simulated tempering.
	In this work, a Monte Carlo step (MCS) is defined by the time unit 
	where $N$ spin trials and a trial of $\beta$ are performed. 
	Figure~\ref{fig:plot-beta} illustrates the time evolution of 
	the relaxation function of $\beta$ defined as
		\begin{align}
		\phi_{\beta}^{(n)}\equiv\frac{\langle\beta^{(n)}\rangle
		-\langle\beta\rangle_{\mathrm{eq}}}{\langle\beta^{(0)}\rangle
		-\langle\beta\rangle_{\mathrm{eq}}},
		\end{align}
	where $\langle\beta^{(n)}\rangle$ denotes the sample average of 
	the inverse temperature after $n$ MCS.
	The expectation with respect to the target distribution in 
	Eq.~(\ref{eq:P_ST}) is denoted by $\langle\cdots\rangle_{\mathrm{eq}}$.
	The initial conditions of $\beta$ are set as $\beta^{(0)}=\beta_R$. 
	The relaxation function monotonically decays to zero with relaxation time. 
	Figure~\ref{fig:plot-beta} indicates that the convergence of $\beta$ with 
	$\delta\not=0$ is more than $10$ times faster than that with $\delta=0$ and 
	that the larger $\delta$ is, the more the relaxation of $\beta$ is 
	accelerated.
	
\begin{figure}[t]
		\includegraphics[width=.95\columnwidth,clip]{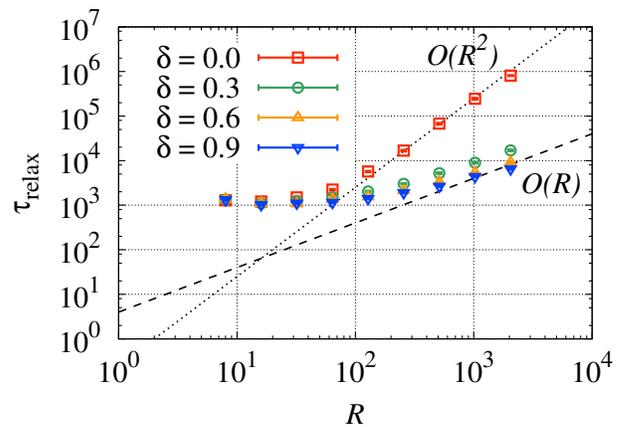}
		\caption{(Color online) 
		$R$-dependence of the relaxation time of $\beta$ 
		in simulated tempering with SDBC.
		The initial state $\bm{S}^{(0)}$ is prepared by 
		the Metropolis-Hastings algorithm at $\beta=\beta_R=0.5$.
		Here, $L=2^5$ and the parameter $\delta$, 
		which characterizes the deviation from DBC, is chosen as 
		$\delta=0.0$ (red square), $\delta=0.3$ (green circle), 
		$\delta=0.6$ (orange triangle), and $\delta=0.9$ (blue inverted triangle).
		The average for each point is taken over $2^{10}$ samples and 
		the error bars are of the order of the symbol sizes. 
		The dotted and dashed line represent asymptotic lines 
		proportional to $R^2$ and $R$, respectively.} 
		\label{fig:plot-taubeta}
		\end{figure}

	In order to evaluate quantitatively the improvement of 
	the relaxation dynamics of the inverse temperature, 
	we measure the relaxation time defined as 
		\begin{align}
		\tau_{\mathrm{relax}}(\epsilon)\equiv
		\inf\{n>0; |\phi_{\beta}^{(n)}|<\epsilon\}. 
		\end{align}
	Figure~\ref{fig:plot-taubeta} represents the $R$-dependence of 
	the relaxation time $\tau_{\mathrm{relax}}(\epsilon)$ with $\epsilon=0.2$.				
	Although no difference between $\delta=0$ and $\delta\not=0$ is observed 
	for small $R$ in Fig.~\ref{fig:plot-taubeta}, the asymptotic behavior of 
	the relaxation time is quite different in each case.
	In the case of $\delta=0$, the relaxation time is asymptotically 
	of order $O(R^2)$, which indicates that the relaxation dynamics of 
	the inverse temperature is diffusive. 
	On the other hand, the relaxation time for $\delta\not=0$ is 
	asymptotically proportional to $R$, which indicates that 
	the relaxation dynamics is ballistic. 
	This difference in the asymptotic behavior is qualitatively consistent 
	with previous works on the one-dimensional simple random 
	walk~\cite{YS2015a,Chen1999,Diaconis2000,Vucelja2014}. 
	This shows that the violation of DBC yields an acceleration of 
	the relaxation of $\beta$.


	\subsection{Empirical transition matrix of inverse temperature}
	
\begin{figure*}[htbp]
		\centering
		\includegraphics[width=.98\textwidth]{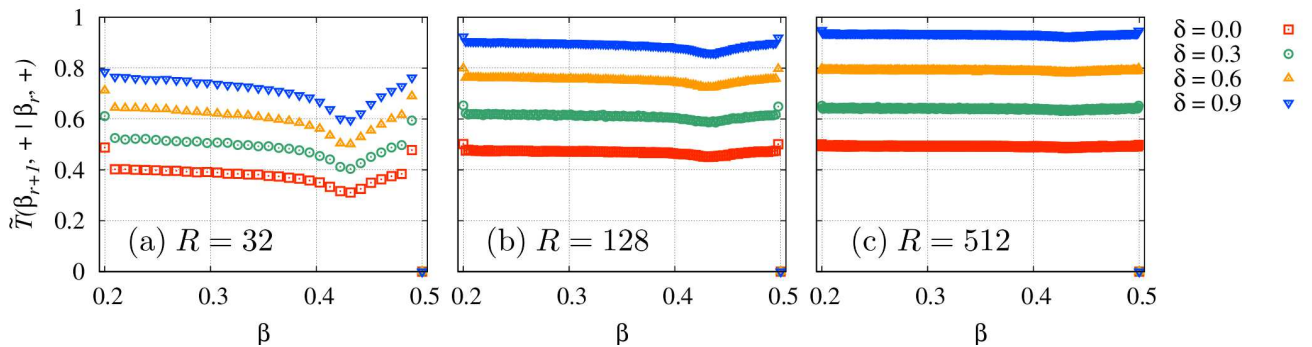}
		\caption{(Color online) 
		$\beta$-dependence of the empirical transition	probability 
		from $(\beta_r,+)$ to $(\beta_{r+1},+)$	with (a) $R=32$ in the left, 
		(b) $R=128$ in the center, and (c)	$R=512$ in the right panel. 
		The linear size of the Ising-spin system is chosen as $L=2^5$ and 
		the parameter $\delta$, which characterizes the deviation from DBC, 
		is chosen as $\delta=0.0$ (red square), $\delta=0.3$ (green circle), 
		$\delta=0.6$ (orange triangle), and $\delta=0.9$ (blue inverted triangle), 
		respectively.}
		\label{fig:a}
		\end{figure*}

\begin{figure*}[htbp]
		\centering
		\includegraphics[width=.98\textwidth]{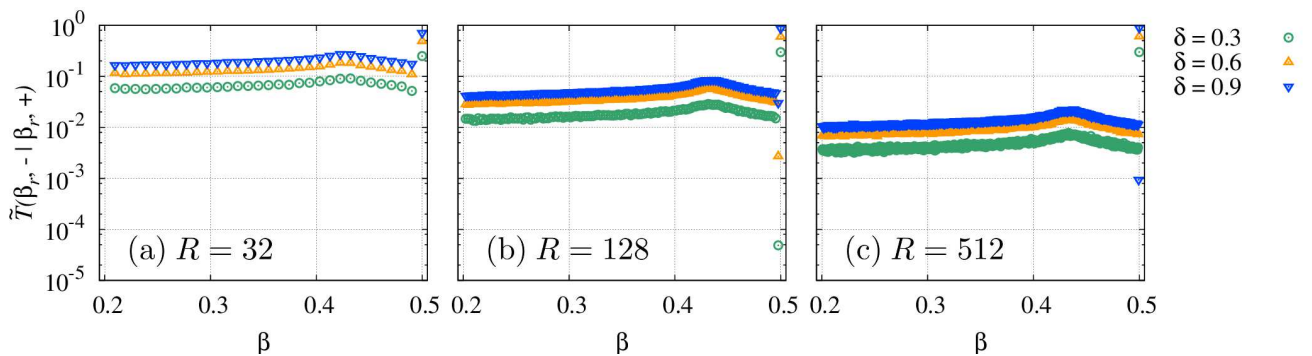}
		\caption{(Color online) 
		$\beta$-dependence of the empirical transition probability 
		from $(\beta_r,+)$ to $(\beta_r,-)$. 
		The left, center, and right panels represent the case (a) $R=32$, 
		(b) $R=128$, and (c) $R=512$, respectively.
		The linear size of the Ising-spin system is chosen as $L=2^5$ and 
		the parameter $\delta$, which characterizes the deviation from DBC, 
		is chosen as $\delta=0.3$ (green circle), 
		$\delta=0.6$ (orange triangle), and 
		$\delta=0.9$ (blue inverted triangle), respectively. 
		Note that the vertical axis is log-scale.}
		\label{fig:b}
		\end{figure*}

	The dynamics of the inverse temperature is explored also 
	through another quantity.
	In implementing the irreversible simulated tempering algorithm,	
	one can easily measure the empirical transition probability 
	with respect to the inverse temperature $\beta$ and 
	the auxiliary random variable $\varepsilon$.
	The empirical transition probability from the state 
	$(\beta_r,\varepsilon)$ to $(\beta_l,\varepsilon')$ is defined as 
		\begin{align}
		&\tilde{T}(\beta_l,\varepsilon'|\beta_r,\varepsilon)
		\notag \\
		&~\equiv
		\frac{\text{\# of transitions from }(\beta_r,\varepsilon) \text{ to } 
		(\beta_l,\varepsilon')}{\text{\# of visit }(\beta_r,\varepsilon)}.
		\end{align}
	In this study, we apply the irreversible simulated tempering algorithm 
	to the two-dimensional Ising model for $2\times 10^5\times R$ MCSs 
	after equilibration and measure the	empirical transition probabilities.
	In our algorithm, explained in the previous section, 
	the non-zero component of the transition probability is 
	$\tilde{T}(\beta_r,\varepsilon|\beta_r,\varepsilon)$, 
	$\tilde{T}(\beta_{r\pm1},\varepsilon|\beta_r,\varepsilon)$, and 
	$\tilde{T}(\beta_r,-\varepsilon|\beta_r,\varepsilon)$ with $\varepsilon=\pm$. 
	Figures~\ref{fig:a} and \ref{fig:b} illustrate the $\beta$-dependence of 
	the empirical transition probabilities.	
	In Fig.~\ref{fig:a}, $\tilde{T}(\beta_{r+1},+|\beta_r,+)$ has a dip 
	around the critical inverse temperature for smaller values of $R$. 
	However, the dip vanishes with increasing number of inverse temperatures and 
	the empirical transition probability becomes flat with respect to $\beta$. 
	Thus, the empirical transition matrix can be approximated 
	by that of the lifted simple random walk in one dimension discussed 
	in Ref.~\cite{YS2015a}. 
	In addition, Fig.~\ref{fig:b} shows that 
	the empirical transition probability for the $\varepsilon$ flip is 
	approximately proportional to $R^{-1}$.
	In Ref.~\cite{YS2015a} it was shown analytically that 
	the lifted simple random walk in one dimension with 
	$O(R^{-1})$ $\varepsilon$-flip probability follows a ballistic process.
	Thus, these results imply the reduction of the convergence rate of 
	the inverse temperature.	
	
\begin{figure}[htbp]
		\centering
		\includegraphics[width=.98\columnwidth]{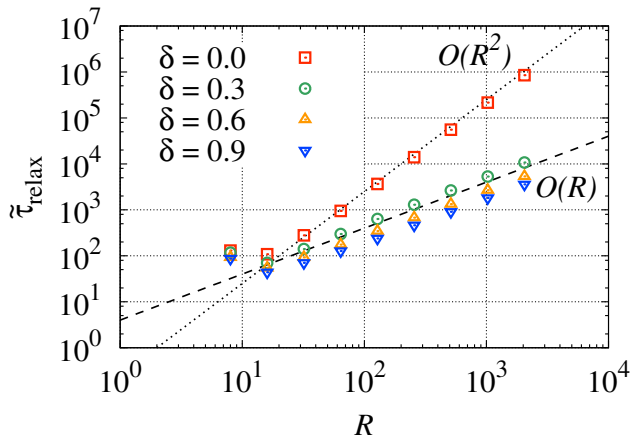}
		\caption{(Color online)
		$R$-dependence of the relaxation time of $\beta$ determined 
		by the empirical transition matrix.
		The linear size of the system is $L=2^5$ and the parameter $\delta$, 
		which characterizes the deviation from DBC, is chosen as 
		$\delta=0.0$ (red square), $\delta=0.3$ (green circle), 
		$\delta=0.6$ (orange triangle), and 
		$\delta=0.9$ (blue inverted triangle), respectively.
		The dotted and dashed lines represent the expected asymptotic 
		form proportional to $R^2$ and $R$, respectively.	
		}
		\label{fig:relaxation-time2}
		\end{figure}

	To evaluate the convergence rate from the empirical transition matrix, 
	we define the relaxation time as follows.
	Let $\lambda_k$ $(k=1,2,...,2R)$ denote the eigenvalues of 
	the empirical transition matrix defined as 
	$\tilde{\mathsf{T}}=\left(\tilde{T}(\beta_l,\varepsilon'|
	\beta_r,\varepsilon)\right)_{1\le r,l\le R; \varepsilon,\varepsilon'=\pm}
	\in\mathbb{R}^{2R\times 2R}$.
	Without loss of generality, the eigenvalues are aligned as 
	$1=\lambda_1>|\lambda_2|\ge\cdots\ge|\lambda_{2R}|$. 
	Then, the relaxation time of the inverse temperature is defined as
		\begin{align}
		\tilde{\tau}_{\rm relax}\equiv
		-\frac1{\ln|\lambda_2|}.
		\end{align}
	Figure~\ref{fig:relaxation-time2} illustrates the $R$-dependence of 
	the relaxation time $\tilde{\tau}_{\rm relax}$ obtained by 
	numerically diagonalizing the empirical transition matrix. 
	As shown in Fig.~\ref{fig:relaxation-time2}, the asymptotic behavior of 
	the relaxation time in both the reversible and irreversible cases is 
	compatible with the results obtained in the previous subsection.
	Thus, the acceleration of the relaxation dynamics of 
	the inverse temperature by the violation of DBC is numerically 
	and theoretically confirmed.	


	\subsection{Autocorrelation function}
		
\begin{figure}[t]
		\includegraphics[width=.95\columnwidth,clip]{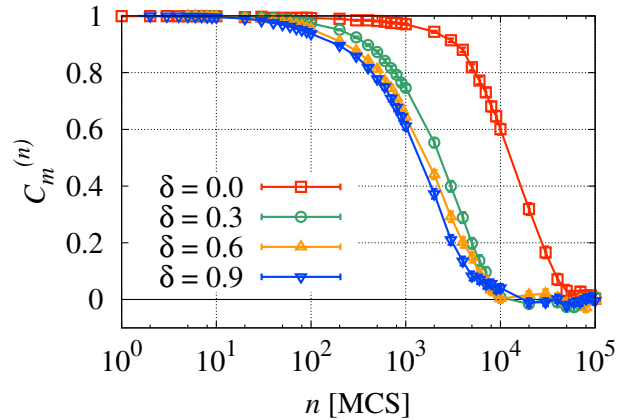}
		\caption{(Color online) 
		Time evolution of the autocorrelation function of the magnetization 
		in the simulated tempering with the SDBC.
		The initial state $\bm{S}^{(0)}$ is prepared by 
		the Metropolis-Hastings algorithm at $\beta=\beta_R=0.5$.
		The linear size of the system and the number of temperature points 
		are chosen as $L=2^5$ and $R=2^9$, respectively.
		The parameter $\delta$, which characterizes the deviation from DBC, 
		is chosen as $\delta=0.0$ (red square), $\delta=0.3$ (green circle), 
		$\delta=0.6$ (orange triangle), and 
		$\delta=0.9$ (blue inverted triangle), respectively. 
		Each data point is averaged over $2^{10}$ samples and 
		the error bars are of the order of the symbol sizes. 
		}
		\label{fig:plot-C}
		\end{figure}

	The acceleration of the relaxation of $\beta$ is expected 
	to promote the acceleration of the relaxation of the magnetization 
	in the Ising model.
	In this subsection, we observe the time evolution of 
	the autocorrelation function of the magnetization whose initial state 
	is prepared as an equilibrium state at $\beta_R$, the lowest temperature 
	in our simulations.
	Let $m=\sum_{i=1}^{N}S_i/N$ denote the averaged spin and 
	let $\langle\cdots\rangle_{\beta}$ be the expectation value 
	with respect to the Gibbs-Boltzmann distribution in Eq.~(\ref{eq:P_beta}).
	Then, we define the (normalized) autocorrelation function of 
	the magnetization in simulated tempering as
		\begin{align}
	 	C_m^{(n)}\equiv\frac{\langle m^{(0)}m^{(n)}\rangle-
	 	\langle m\rangle_{\beta_R}\langle m\rangle_{\rm eq}}{
	 	\langle m^2\rangle_{\beta_R}-
	 	\langle m\rangle_{\beta_R}\langle m\rangle_{\rm eq}},
	\end{align}
	where $\langle m^{(0)}m^{(n)}\rangle$ denotes the sample average of 
	the correlation between the initial averaged spin and that after $n$ MCSs.
	Figure~\ref{fig:plot-C} illustrates the time evolution of $C_m^{(n)}$, 
	which indicates that the violation of DBC reduces the relaxation rate 
	in the autocorrelation function by a factor as large as ten (10).
	This reduction is affected by the acceleration of the relaxation of $\beta$.
	
\begin{figure}[t]
		\includegraphics[width=.95\columnwidth,clip]{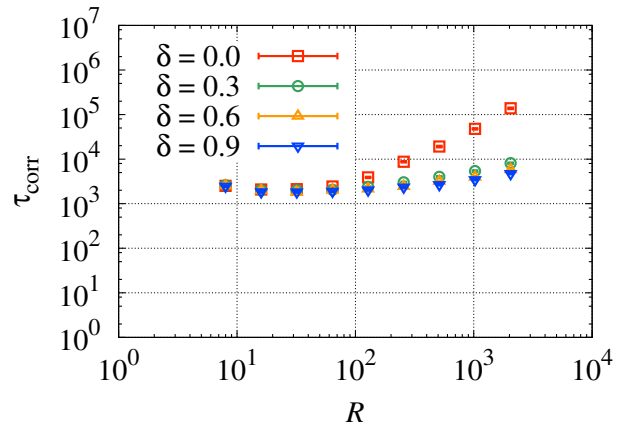}
		\caption{(Color online) 
		$R$-dependence of the autocorrelation time in simulated tempering 
		with SDBC.
		The initial state $\bm{S}^{(0)}$ is prepared by 
		the Metropolis-Hastings algorithm at $\beta=\beta_R=0.5$.
		The value $L=2^5$ is used and the parameter $\delta$, 
		which characterizes the deviation from DBC, is chosen as
		$\delta=0.0$ (red square), $\delta=0.3$ (green circle),	
		$\delta=0.6$ (orange triangle), and 
		$\delta=0.9$ (blue inverted triangle), respectively. 
		Each data point is averaged over $2^{10}$ samples and 
		the error bars are of the order of the symbol sizes. 
		}
		\label{fig:plot-tau}
		\end{figure}

	In order to evaluate quantitatively the improvement of 
	the relaxation dynamics of the autocorrelation, the autocorrelation time 
	is defined as
		\begin{align}
		\tau_{\mathrm{corr}}(\epsilon)\equiv\inf(n>0; |C_m^{(n)}|<\epsilon),
		\end{align}
	and especially $\tau_{\mathrm{corr}}=\tau_{\mathrm{corr}}(\epsilon=0.2)$.
	Figure~\ref{fig:plot-tau} represents the $R$-dependence of 
	the autocorrelation time $\tau_{\mathrm{corr}}$.
	As shown in Fig.~\ref{fig:plot-tau}, while there is no difference 
	observed between $\delta=0$ and $\delta\not=0$ for small $R$, 
	the autocorrelation time is improved for relatively large $R$.
	The above results confirm that the violation of DBC improves 
	the efficiency of the simulated tempering algorithm with respect to 
	the sampling of both $\beta$ and $\bm{X}$.


	\section{\label{Sect:Summary}Summary and Discussion}
	
	We have constructed an irreversible simulated tempering algorithm 
	by introducing the lifting technique based on the methodology of SDBC 
	to the update scheme of the inverse temperature.
	Benchmarks for the Ising model show that our algorithm accelerates 
	the relaxation dynamics of inverse temperature and 
	the autocorrelation function of the magnetization compared to 
	the traditional simulated tempering algorithm based on DBC.
	These results show that the lifting technique can improve 
	the efficiency of extended-ensemble methods.
	Furthermore, we consider the empirical transition probability 
	with respect to the inverse temperature and the lifting parameter 
	to investigate the relaxation dynamics of the inverse temperature in detail.
	It is easily measured during numerical simulations 
	in the irreversible simulated tempering algorithm.
	We found that the empirical transition matrix is approximately the same as 
	the transition matrix of the lifted simple random walk in one dimension 
	discussed in Ref.~\cite{YS2015a}. 
	Thus, it is theoretically confirmed that the lifting technique accelerates 
	the relaxation dynamics of the inverse temperature.
	
	Although we used our proposed algorithm for the Ising model 
	in two dimensions in this paper, our algorithm is, 
	in principle, applicable to any other system such as Potts model, 
	Heisenberg spin glass, and protein systems.
	It is also possible to combine other update schemes of 
	the configuration of target systems, such as the Swendsen-Wang algorithm 
	and the Wolff algorithm, instead of the Metropolis-Hastings algorithm.
	Our algorithm could take over these advantages from 
	the traditional simulated tempering method.
	It is worth investigating whether the irreversible simulated tempering 
	combined with such an update scheme works effectively 
	in a system with a first-order phase transition and spin glasses.
	
	In this study, all inverse temperatures were arranged at equal distances 
	and the weight factor $g_r$ was estimated by an exact numerical method.
	The choice of the set of inverse temperatures $\{\beta_r\}$ and 
	parameters $g_r$ affect the efficiency of simulated tempering.
	Several studies have proposed their efficient 
	choices~\cite{Hansmann1997, Mitsutake2000, Park2007, Valentim2014}.
	A promising way for 
	estimating the weight factor 
	is to implement the irreversible simulated tempering 
	algorithm which is our current work in progress~\cite{YS2016b}. 
		

		\begin{acknowledgments}
		The authors are grateful to S.~Todo for useful comments	and 
		for bringing the method of Ref.~\cite{Kastening2001} to our notice. 
		Y.S. is supported by a Grant-in-Aid from 
		the Japan Society for Promotion of Science (JSPS) Fellows 
		(Grant No. 26$\cdot$7868).
		K.H. is supported by Grants-in-Aid for Scientific Research from MEXT, 
		Japan (Grant Nos.\ 25610102 and 25120010), 
		and JSPS Core-to-Core program 
		``Nonequilibrium dynamics of soft matter and information.''
		\end{acknowledgments}



\end{document}